\documentclass[amsmath,prl,aps,twocolumn]{revtex4}
\usepackage{amsthm,amsfonts,graphicx,verbatim}

\newcommand{\Tr}{{\rm Tr}}
\newcommand{\tr}{{\rm tr}\/}

\newcommand{\D}{{\rm d}}

\newcommand{\be}{\begin{equation}}
\newcommand{\ee}{\end{equation}}
\newcommand{\bea}{\begin{eqnarray}}
\newcommand{\eea}{\end{eqnarray}}

\newcommand{\la}{\langle}
\newcommand{\ra}{\rangle}
\newcommand{\lb}{\left[}
\newcommand{\rb}{\right]}
\newcommand{\lp}{\left(}
\newcommand{\rp}{\right)}

\renewcommand{\epsilon}{\varepsilon}

\begin{document}
\title{Quantum Noise as an Entanglement Meter}
\author{Israel Klich$^{1}$ and Leonid Levitov$^{1,2}$ }
\affiliation{${}^1$ Kavli Institute for Theoretical Physics,
University of California Santa Barbara, CA
93106 \\
${}^2$ Department of Physics, Massachusetts Institute of Technology, Cambridge MA 02139}


\begin{abstract}
Entanglement entropy, which is a measure of quantum correlations between separate parts of
a many-body system, has emerged recently as a fundamental quantity in broad areas of theoretical
physics, from cosmology and field theory to condensed matter theory and quantum information. The universal appeal of the entanglement entropy concept is related, in part, to the fact that it is defined solely in terms of  the many-body density matrix of the
system, with no relation to any particular observables. However, for the same reason, it has not been clear how to access  this  quantity experimentally. Here we
derive a universal relation between entanglement entropy and the fluctuations of current flowing
through a quantum point contact (QPC) which opens a way to perform a direct
measurement of entanglement entropy. In particular, by utilizing space-time duality
of 1d systems, we relate electric noise generated
by opening and closing the QPC periodically in time with the seminal ${\cal S}=\frac13 \log L$ prediction of conformal field theory. 
%
\end{abstract}

 \maketitle

Recent years have witnessed a burst of interest in the phenomena of quantum entanglement, and in particular, in entanglement entropy, a fundamental characteristic describing quantum many-body
correlations between two parts of a quantum system.
This quantity first emerged in field theory and cosmology~\cite{Bombelli86,Holzhey94} under the name of ``geometric entropy,'' and subsequently was adopted by quantum information theory. The notion of entanglement entropy has provided a framework for analyzing quantum critical phenomena \cite{VidalLatorreetal,Refael Moore,Cardy} and quantum quenches~\cite{CalabreseCardyTimeDependentEnt,Bravyi06,EisertOsborne,Cramer08}. Recently it was used as a probe of complexity of topologically ordered states~\cite{Kitaev Preskill,Levin Wen,Fradkin}.
In addition, this quantity is of fundamental interest for quantum information theory as a measure of the resources available for quantum computation~\cite{Bennett96} as well
as for numerical approaches to strongly correlated systems~\cite{Verstraete04}.

\begin{figure}
\includegraphics*[width=2.8in]{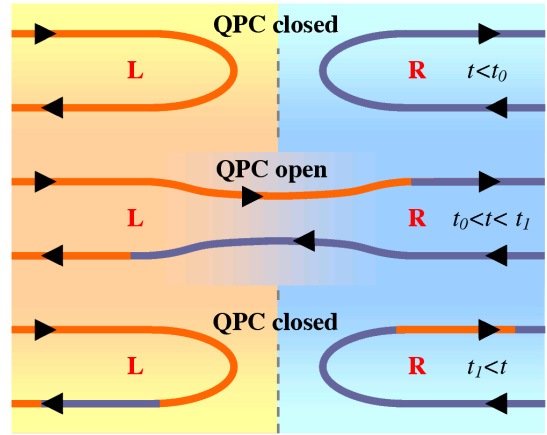}
\caption{
Schematic of a quantum point contact (QPC) with transmission changing in time.
The left and right leads are initially
disconnected, then connected at $t_0<t<t_1$, and then disconnected again.
Electron transport, taking place at $t_0<t<t_1$
makes electrons delocalized among the two leads, generating entanglement and current fluctuations.} 
\label{fig1}
\end{figure}

Can the entanglement entropy be measured? 
Here we identify a system where such a measurement is possible, thereby offering an affirmative answer to this question.
In particular, we establish a relation between the entropy and quantum noise in a quantum point contact (QPC) \cite{van Wees}, an electron beam-splitter with tunable transmission and reflection. In essense, the QPC serves as a door between electron reservoirs, which can be opened and closed on demand (see Fig.\ref{fig1}). 
We show that the fluctuations of electric current flowing through the QPC can be used to quantify the entanglement generated by the connection, and thereby measure the entanglement entropy. 

On some level, the very idea of measuring a quantity that encodes information about many-body correlations of a large number of particles, which is what the entanglement entropy is, may seem totally bizarre. Yet, as we shall see, the situation with the entanglement entropy is different from, for example, the many-body density matrix that depends on coordinates of all particles in the system and is thus indeed very difficult to measure. In the free fermion QPC problem analyzed below, all multi-particle correlations in the Fermi sea that are relevant for entanglement are fully accounted for by temporal correlations of current flowing through the QPC. As a result, perhaps somewhat surprisingly, noise measurement provides sufficient information needed to determine the entanglement entropy. 

The situation considered here, which involves connecting and then disconnecting
two parts of the system,
is in a sense dual to the more conventional picture used in Refs.\cite{Holzhey94,VidalLatorreetal,Refael Moore,Cardy}. In the latter approach, the many-body ground state of a translationally invariant system is analyzed using a finite region in space. In our case, a window in time is used, 
$t_0<t<t_1$, during which particles can delocalize among the reservoirs, making them entangled.

The relation between entanglement and noise is at the heart of recent proposals~\cite{Beenakker03,Samuelsson03} to use 
current partitioning by scattering on a QPC for producing entangled particle pairs. Ref.\cite{KlichRefaelSilva} considered measuring the entanglement entropy, emphasizing its relation
with particle number statistics, however without making connection to electron transport and current fluctuations.
We also note that 
generation of entanglement  
was recently analyzed
for critical Hamiltonians~\cite{CalabreseCardyTimeDependentEnt}, for generic Hamiltonians~\cite{Bravyi06,EisertOsborne}, as well as in 
a QPC under bias voltage \cite{Beenakker}.


Our analysis reveals a relation between entanglement production in a driven QPC 
with the theory of quantum noise~\cite{LevitovLesovik}, describing quantum and classical current fluctuations in this system. This approach, known as Full Counting Statistics (FCS), describes
the probability distribution of transmitted charge using 
the generating function $\chi(\lambda)=\sum_{n=-\infty}^\infty
P_ne^{i\lambda n}$, where $P_n$ is the probability to transmit $n$ charges in total. The function $\chi(\lambda)$ encodes all FCS higher moments, or cumulants $C_m$: 
\be\label{Def of chi}
  \log\chi(\lambda)=\sum_{m=1}^\infty {(i\lambda)^mC_m\over m!}
,
\ee
where $C_1$, $C_2$, $C_3$... describe properties of the distribution $P_n$ such as the mean $\bar n$, the variance $\la (n-\bar n)^2\ra$, the skewness $\la (n-\bar n)^3\ra$ , etc. The cumulant $C_2$ is available from routine noise measurement. 
Recently, $C_3$ has been measured in tunnel junctions~\cite{Reulet03,BomzeReznikov} and in QPC~\cite{Gershon07}, while
cumulants up to 5th order where measured in quantum dots~\cite{Fujisawa,Gustavsson}. 

Below we establish a universal relation between FCS (\ref{Def of chi}) and the entanglement entropy generated 
in the process shown in Fig.\ref{fig1}. Only even cumulants are shown to contribute to entropy:
\begin{eqnarray}\label{Entropy from cumulants}
{\cal S}=\sum_{m>0} {\alpha_m\over  m !}C_{m} 
,\quad
\alpha_m=
\Big\{
\begin{array}{cc}
    (2\pi)^m |B_{m}|, &  $m$\,\,{\rm even} \\
  0,  &  $m$\,\,{\rm odd} 
\end{array}
,
\end{eqnarray}
where $B_m$ are Bernoulli numbers~\cite{Bernoulli_numbers} ($B_2=\frac16$, $B_4=-\frac1{30}$, $B_6=\frac1{42}$...). The first few contributions are:
\begin{eqnarray}\label{Entropy C2 C4 C6}
  {\cal S}={\pi^2\over 3} C_2+{\pi^4\over 15}C_4+{2\pi^6\over 945}C_6+...
\end{eqnarray}
This relation, which is completely general and valid for arbitrary driving, can be used to determine the entanglement entropy from measured values of FCS moments. 
Similar relation can be derived for other quantities of interest, such as Renyi entropies, or single copy entropy \cite{Eisert Cramer}.

In particular, quantum noise generated in the QPC switching on and off (Fig.\ref{fig1}) directly corresponds to the entanglement entropy found for conformal field theory, $S=\frac c3 \log L$, where $L$ is the size of  window in  space and $c $ the conformal charge~\cite{Holzhey94,VidalLatorreetal,Cardy}. In this case the current fluctuations are gaussian ($C_{m\neq 2}=0$), with a logarithmic variance 
$C_2={1\over \pi^2}\log\frac{t_1-t_0}{\tau}$, 
where $\tau$ is a short time cutoff set by the QPC switching rapidity. Combined with Eq.(\ref{Entropy C2 C4 C6}) this gives entropy $S\sim\frac13\log |t_1-t_0|$. Below we discuss how this logarithmic dependence can be verified using the setup shown in Fig.\ref{fig1}.

Entanglement entropy is conventionally defined as the von Neumann entropy ${\cal S}(\rho)=-{\bf Tr} \rho\log \rho$, where $\rho$ is the reduced density matrix of a pure quantum state, made ``impure'' by confining it to a certain space region \cite{Bombelli86,Holzhey94}.
In our case, the many-body state evolves as a pure state while the QPC is open
(see Fig.\ref{fig1}), after which the reduced density matrix of the lead $L$ is given by
\begin{eqnarray}
  \rho_L(t_1)={\bf Tr}_R ({\bf U}(t_1,t_0)\rho_0{\bf U}^{\dag}(t_0,t_1))
.
\end{eqnarray}
Here $\rho_0$ is the initial density matrix of the system,
${\bf U}$ is the many-body evolution between $t_0$ and $t_1$, and ${\bf Tr}_{R}$ is a partial trace over degrees of
freedom in the lead $R$. 

Entropy production in the lead $L$ as a result of QPC opening and closing is described by
\be\label{eq:DeltaS}
\Delta {\cal S} = {\cal S}(\rho_L(t))-{\cal S}(\rho_0)
\ee
where the last term accounts for the entropy in the initial state. Because at finite temperature both terms in Eq.(\ref{eq:DeltaS}) are proportional to the lead volume, they can be large for macroscopic leads. The increment $\Delta  {\cal S}$, however, remains well defined regardless of the lead volume. 

Below we focus on the zero temperature case when $\rho_0$ is a pure state,
described as a filled Fermi sea in the full system $L+R$, in which case the second term in Eq.(\ref{eq:DeltaS}) vanishes, giving $\Delta  {\cal S}={\cal S}(\rho_L(t))$.
We associate with $\rho_0$ a Fermi projection operator $n$
in the single-particle space $\la
E|n|E'\ra=\delta_{E,E'}\theta(E_F-E)$, where $E_F$ is the Fermi energy. The evolved system is described by a rotated Fermi projection 
$n_U=UnU^{\dag}$, where $U$ is the
unitary evolution of the single-particle modes.

Our first step will be to express the entropy in terms of single-particle quantities. 
For a generic gaussian state, Wick's theorem for operator products is satisfied  in $L+R$, and
therefore in particular it holds in $L$~\cite{Peschel}. Therefore the 
reduced density matrix $\rho_L$ is also gaussian:
\be\label{eq:gaussian_state}
\rho_L={1\over Z} e^{-\tilde{H}_{ij}a^{\dag}_ia_j}
\quad (i,j\ {\rm in}\ L)
\ee
for some $\tilde{H}$. We define a single-particle quantity 
$m_{ij}= {\bf Tr}\,  \rho_L a^{\dag}_ia_j$.
For the evolved system, described by 
$n_U$, Wick's theorem gives $m_{ij}=(n_U)_{ij}$.
In what follows it will be convenient to extend $m$ to $L+R$ 
by setting 
\be\label{eq:M_definition}
M=P_L n_U P_L
, 
\ee
with $P_L$ a projection on
the modes in $L$, so that $M=m$ in $L$ and $M=0$ in $R$.

Entropy can be expressed through $m_{ij}$ for a generic gaussian state (\ref{eq:gaussian_state}).
Because of Fermi-Dirac statistics, $m=(1+e^{\tilde{H}})^{-1}$,
which gives
$\tilde{H}=\log(m^{-1}-1)$ \cite{Peschel}. Extending $m$ to $M$ in $L+R$, we write
the entropy as
\be\label{EntropyM}
{\cal S}(\rho)=-\Tr \lb M\log M+(1-M)\log(1-M)\rb
\ee
where now the trace is taken in
the space of single-particle modes in $L$.  

Transport in a QPC is described by time-dependent transmission and reflection amplitudes $A(t)$, $B(t)$. In a Schr\"odinger representation, the scattering states are
\be\nonumber
U\begin{bmatrix}
 |x\ra_L  \\
 |x\ra_R \\
\end{bmatrix}
=
\begin{bmatrix}
  B(t_x) & A(t_x) \\
  -\bar A(t_x) & \bar B(t_x) \\
\end{bmatrix}
\begin{bmatrix}
 |x(t)\ra_L  \\
 |x(t)\ra_R \\
\end{bmatrix}
,\quad
x<0<x(t)
,
\ee
and $U |x\ra_{L,R}=|x\ra_{L,R}$ otherwise.
Here $x(t)=x+v_Ft$,
$t_x=-x/v_F$ is the time of arrival at the scatterer, $v_F$ is the Fermi velocity,
and $|x\ra_{L,R}$ describes incoming ($x<0$) and outgoing ($x>0$) wavepacket states in the leads.

In FCS approach it is convenient to work in a time representation~\cite{LevitovLesovik04}, labeling states by times of arrival at the scatterer $t_x$. In this representation 
the initial Fermi projection is given
by $n(t,t')={1\over 2\pi i(t-t'+i0)}I$ with $I$
 a $2\times 2$ identity matrix (in $L$, $R$). The evolved state $n_U$ is
\begin{eqnarray}\nonumber
n_U(t,t')=U(t)n(t,t')U^{\dag}(t')
,\quad
U(t)=
\begin{bmatrix}
  B(t) & A(t) \\
  -\bar A(t) & \bar B(t) \\
\end{bmatrix}
.
\end{eqnarray}
These relations are inessential for the derivation of our main result, they serve here illustration purpose only.

Our next step is to relate the quantity $M$, Eq.(\ref{eq:M_definition}), 
and the FCS generating function \eqref{Def of chi} which can be expressed as a functional determinant~\cite{LevitovLesovik}:
\be\label{FCS}
   \chi(\lambda)=
 \det(1-n +n  U^{\dag}e^{i\lambda P_L}U e^{-i\lambda
  P_L})
.
\ee
%
This determinant must be 
properly regularized for infinitely deep Fermi sea \cite{MuzukantskiiAdamov,Avron
Graf}. 
For our purposes we proceed to treat it
as a finite matrix although a more rigorous treatment using $C^\ast$ algebra techniques is possible \cite{Avron Graf}. In the spirit of \cite{Avron Graf,MuzukantskiiAdamov}, we rewrite \eqref{FCS} as~\cite{Abanov Ivanov}:
\begin{eqnarray}&
\det(e^{i\lambda P_L n}(1-n +ne^{-i\lambda
  P_L} U^{\dag}e^{i\lambda P_L}U)e^{-i\lambda P_L n}).
\end{eqnarray}
Using the identities $e^{i \lambda P_L n}(1-n)=1- n$ and $e^{i \lambda P_L (1-n)}n=n$ we find:
\begin{eqnarray}&
\chi(\lambda)=\det((1-n + n U^{\dag}e^{i\lambda P_L}U  )e^{-i\lambda P_L n}).
\end{eqnarray}
Next, we insert $U^{\dag}U$ in the determinant to obtain
\begin{eqnarray}&
\chi(\lambda)=\det(U(1-n + n U^{\dag}e^{i\lambda P_L}U  )e^{-i\lambda P_L n}U^{\dag})\\ \nonumber &
=\det((1-n_U + n_U  e^{i\lambda P_L} )e^{-i\lambda (P_L n)_U})\\ \nonumber &
=\det((1+n_U P_L(e^{i\lambda} -1) )e^{-i\lambda (P_L n)_U})
,
\end{eqnarray}
where in the last line we used: $e^{i\lambda P_L}=1+(e^{i\lambda}-1)  P_L$.
Finally, noting that $\det(1+AP_L)=\det(1+P_LAP_L)$ for any matrix $A$, we arrive at
\be\label{eq:chi-M}
\chi (\lambda)=\det\Big((1-M +M 
e^{i\lambda })e^{-i\lambda (n P_L)_U}\Big)
,
\ee
where $M$ is the quantity (\ref{eq:M_definition}) which defines the entropy.

Now, with the help of the relation (\ref{eq:chi-M}) we can express the spectral density of $M$, Eq.(\ref{eq:M_definition}), which lies between $0$ and $1$, through $\chi(\lambda)$. Indeed, changing parameter $\lambda$ to $z=(1-e^{i\lambda})^{-1}$ yields $z-M$ under the determinant: $\chi (z)=\det\lp (z-M) e^{-i(n P_L)_U\lambda(z)}(1-e^{i\lambda(z)})\rp$. From that we can write the spectral density of $M$ as
\be\label{eq:mu(z)-chi(z)}
\mu(z)={1\over \pi}{\rm Im}\,\partial_z\log\chi(z-i0)+A\delta(z)+B\delta(z-1)
,
\ee
where the coefficients $A$ and $B$ depend on $\rm\dim{M}$, $\Tr (nP_L)_U$, and $C_1$. Hereafter we ignore the delta function terms because $z=0,1$ do not contribute to the expression (\ref{EntropyM}) for the entropy which we rewrite as
\be\label{S from mu}
  {\cal S}=-\int_0^1\D z\mu(z)\lb z\log z+(1-z)\log(1-z)\rb
\ee
Now it is straightforward to evaluate ${\cal S}$ by substituting (\ref{Def of chi}) into
\eqref{eq:mu(z)-chi(z)},\eqref{S from mu}
with $\lambda(z)=\pi-i\log\lp \frac1{z}-1\rp$,
and integrating by parts over $z$. We obtain series
${\cal S}=\sum_{m=1}^{\infty} {\alpha_{ m}\over m!}C_{ m}$,
%
%
where the coefficients $\alpha_m$, after changing variable to $u={1\over
2}\log({z\over 1-z})$, take the form
\begin{eqnarray}
\alpha_m={(-2)^m\over\pi}\int_{-\infty}^{\infty}\D u {u\over
\cosh^2{u}}\,{\rm Im}\,\lp {i\pi\over 2}+u\rp^m
\end{eqnarray}
Finally, for $m\geq 2$, shifting the contour of integration as $u\to u-i\frac{\pi}2$, 
%
%
yields~\cite{GradshteynRyzhik} our main results (\ref{Entropy from cumulants}) and (\ref{Entropy C2 C4 C6}).

In order to compute FCS, it is convenient to return to the expression (\ref{FCS}) and apply the Riemann-Hilbert (RH) method introduced in
\cite{MuzukantskiiAdamov}. In this approach, one must factor the time-dependent matrix
$R(t)=U^{\dag}(t)e^{i\lambda P_L}U(t)e^{-i\lambda P_L}$ in (\ref{FCS}), i.e. find
matrix valued functions $X_{\pm}(z)$, analytic in  the upper/lower half plane of complex $z$, respectively, such that on
the real line:
\begin{eqnarray}
X_{+}(t)=X_{-}(t)R(t)
\end{eqnarray}
with normalization
$X_{\pm}(z)\rightarrow I$ at $|z|\rightarrow\infty$.

We consider QPC switching between the on and off states several
times $t^{(1)}_0<t^{(1)}_1<...<t^{(N)}_0<t^{(N)}_1$. The RH problem is solvable in the case of abrupt switching because
$R$ commutes with itself at different times: $R=I$ 
in the off state, $R=e^{-i\lambda (P_R-P_L)}$
in the on state. 
The solution of the RH problem is then given by the functions
\begin{eqnarray}\label{RH solution}
X_{\pm}(z)=\exp\lp {\lambda\over 2\pi} (P_R-P_L) \sum_{i=1}^N \log {z-t^{(i)}_{0}\pm i0\over z-t^{(i)}_{1}\pm i0}\rp 
.
\end{eqnarray}
To find the determinant (\ref{FCS}) with these $X_{\pm}$ we use the RH method \cite{MuzukantskiiAdamov} to evaluate the derivative of $\log\chi$:
\begin{eqnarray} \nonumber
&&\partial_{\lambda}\log\chi(\lambda)= \int  \tr
 \lp {1\over 2\pi i} X_{+}^{-1}\partial_t X_{-}\partial_{\lambda} R 
 \rp \D t
=-\frac{\lambda}{2\pi^2}G
,
\\\label{eq:G_N}
&& G =\sum_{i, j=1}^N\log\frac{t_1^{(i)}-t_0^{(j)}}{t_0^{(i)}-t_0^{(j)}}
+\log\frac{t_1^{(i)}-t_0^{(j)}}{t_1^{(i)}-t_1^{(j)}}
,
\end{eqnarray}
where for $i=j$ the denominators must be replaced by a short-time cutoff $\tau$.
This gives gaussian charge statistics
\be
\chi(\lambda)=e^{-\frac12\lambda^2 C_2}
,\quad
C_2=\frac1{2\pi^2}G
.
\ee
Because in this case the only nonvanishing cumulant is $C_2$, we have ${\cal S}={\pi^2\over 3}C_2$, which gives ${\cal S}={1\over 3}\log{t_1-t_0\over \tau}$ for a single QPC switching, in agreement with the ${\cal S}\sim\frac13 \log L$ relation \cite{Holzhey94}. 
The case of multiple switching provides realization of the situation studied in Ref.\cite{Cardy}.


\begin{figure}
\includegraphics*[width=3.4in]{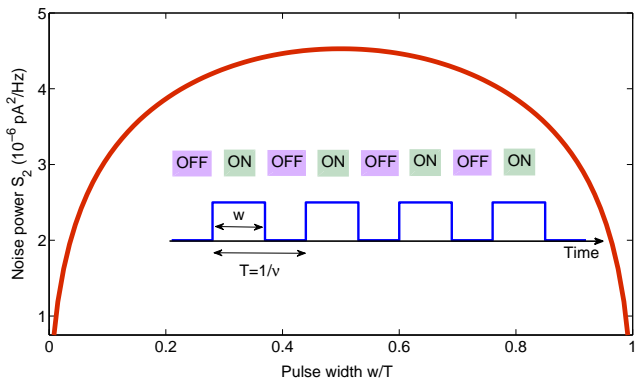}
\caption{Entanglement entropy and current fluctuations. Noise power (\ref{eq:S2})
in a QPC driven by a pulse train, plotted as a function of the pulse width. 
Parameters used: driving frequency $\nu=500\,{\rm MHz}$, short-time cutoff $\tau=20\,{\rm ps}$. The noise as well as the entropy production are symmetric under $w\to T-w$.
Note that at a narrow pulsewidth $w\ll T$ the dependence (\ref{eq:S2}) reproduces the $\frac13\log L$~\cite{Holzhey94} behavior of the entropy.
} 
\label{fig2}
\end{figure}

\begin{figure}
\includegraphics*[width=3.4in]{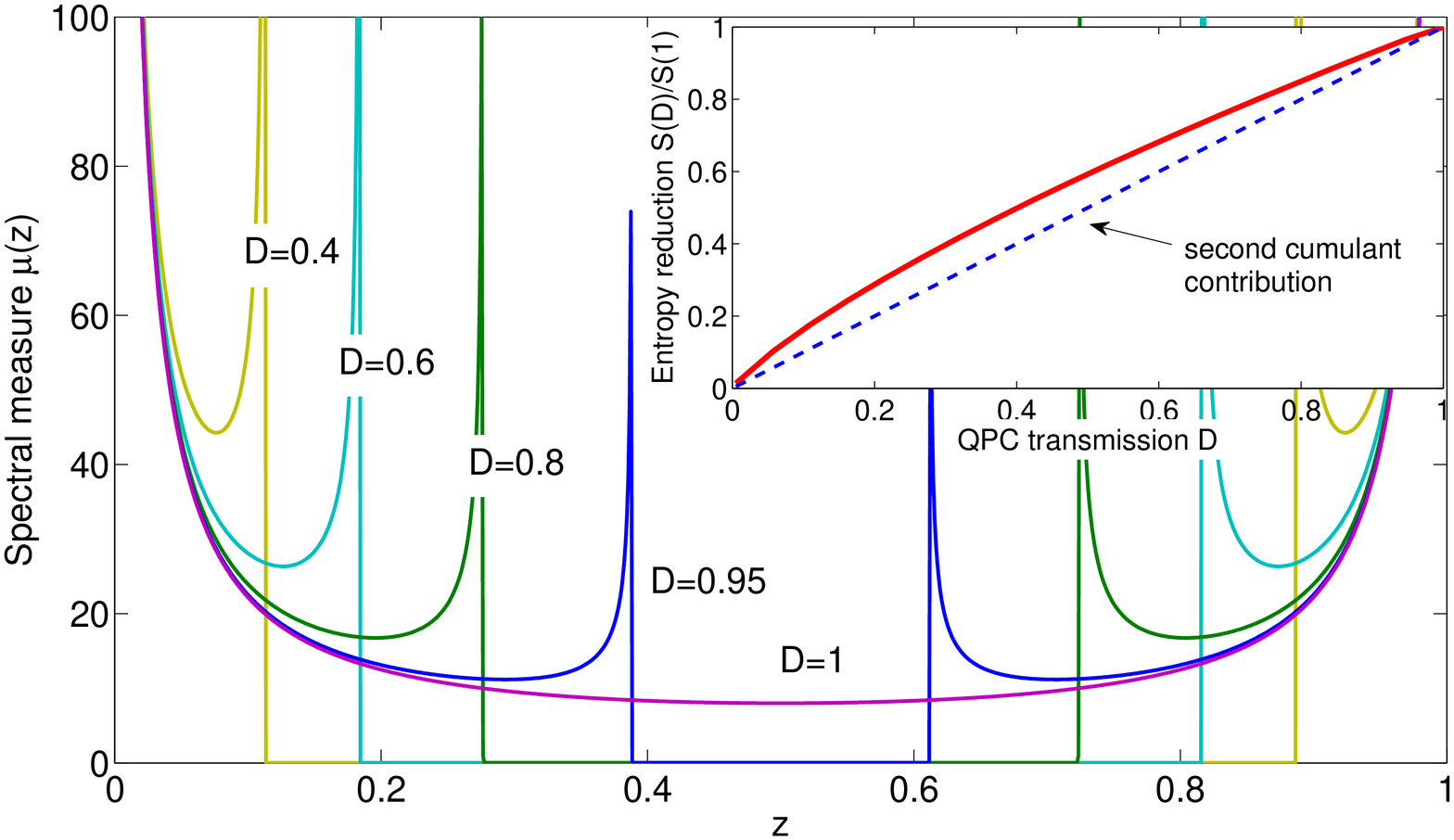}
\caption{The effect of imperfect transmission in QPC. Spectral measure (\ref{eq:mu(z)_D}) scaled by $G/2\pi^2$, for a QPC driven by pulses with transmission $D$ in the ``on'' state (see Fig.\ref{fig2}). At $D<1$ the entropy (\ref{S from mu})  
is reduced by a constant factor (inset) with
the ${\cal S}\sim\log\sin\pi\nu w$ dependence unchanged. 
} 
\label{fig3}
\end{figure}

These predictions can be tested by measuring noise in a QPC driven by a periodic train of pulses (see Fig. 2). For $N$ identical pulses, the relation (\ref{eq:G_N}) at large $N$ yields
\be\label{eq:C_2=log sin}
C_2(N)\approx \frac{N}{\pi^2}\log\frac{\sin\pi\nu w }{\pi\nu\tau}
,\quad
\nu=1/T
.
\ee
%
Thus in a periodically driven QPC there is a finite entropy production per cycle at a rate $\D {\cal S}/\D t=\frac13\nu \log\frac{\sin\pi\nu w}{\pi\nu\tau}$. Fluctuations (\ref{eq:C_2=log sin}) with $C_2\propto N$ 
correspond to electric noise with spectral power
%
\be\label{eq:S2}
S_2 = \frac{e^2\nu}{\pi^2}\log\frac{\sin\pi\nu w}{\pi\nu\tau}
\ee
at frequencies below $\nu$. For a short pulsewidth $w$, the dependence \eqref{eq:S2} becomes 
$S_2=\frac{e^2\nu}{\pi^2}\log\frac{w}{\tau}$, identical to the entropy
for a single pulse.

The result (\ref{eq:S2}) must be compared with thermal noise. At a driving frequency $\nu=500\,{\rm MHz}$, the effective temperature corresponding to (\ref{eq:S2}) is $T_{\rm eff}=\frac{h\nu}{\pi^2k_{\rm B}}\log\frac{\sin\pi\nu w}{\pi\nu\tau}\approx 25\,{\rm mK}$. 
In practice it may be possible to relax the constraint due to small $T_{\rm eff}$ by detecting the noise (\ref{eq:S2}) at frequencies somewhat higher then $k_BT_{\rm eff}/h$, detuned from the thermal noise spectral window. 

How sensitive are these results to imperfections in QPC transmission? 
It is straightforward to incorporate transmission $D=|A|^2<1$ in the ``on'' state in the RH analysis because the matrices $R(t)$ in (\ref{FCS}) still commute at different times. Instead of $e^{\pm i\lambda}$, the eigenvalues of the $R$ matrix are now $e^{\pm i\lambda_\ast}$ with $\sin\frac12\lambda_\ast=\sqrt{D}\sin\frac12\lambda$ \cite{LevitovLesovik04}. Making this change, we obtain
\be\label{eq:chi_eta(lambda)}
\chi(\lambda)=e^{-\frac{\lambda^2_\ast}{4\pi^2}G}
,
\ee
with $G$ given by (\ref{eq:G_N}) as above. Because this $\chi(\lambda)$ is non-gaussian, with nonzero higher cumulants, the simplest way to find the entropy is to use its relation with the spectral density of $M$, Eqs.(\ref{eq:mu(z)-chi(z)}),(\ref{S from mu}). Using (\ref{eq:chi_eta(lambda)}) along with the relations between $\lambda_\ast$, $\lambda$ and $z$, we find
%
\be\label{eq:mu(z)_D}
\mu(z)=\frac{G}{\pi^2}\frac{D}{z(1-z)}\,{\rm Re}\,\frac{|1-2z|}{\sqrt{D^2-4D z(1-z)}}
.
\ee
As illustrated in Fig.3, at $D<1$ the function $\mu(z)$ vanishes in the interval $z_-<z<z_+$, $z_{\pm}=\frac12\lp 1\pm\sqrt{1-D}\rp$. 

The entropy, found from 
\eqref{S from mu} and \eqref{eq:mu(z)_D}, has the same logarithmic dependence (\ref{eq:G_N}) on the times $t_{0,1}^{(i)}$ as above, albeit with a $D$-dependent prefactor.
Thus, the predicted dependence ${\cal S}\sim\log\sin\pi\nu w$ remains robust. The behavior of the rescaling factor $F={\cal S}(D)/{\cal S}(1)$ (Fig.3 inset) indicates that
entropy reduction due to imperfect transmission in QPC can be attributed mostly to the change in the second cumulant, $C_2=\frac{D}{2\pi^2}G$, with a relatively small correction due to higher cumulants.

From the quantum information perspective, it is important to isolate the part of the entropy accessible to local operations (i.e. respecting particle conservation in each lead) \cite{Beenakker}. The particle-number restricted entropy may be easily obtained from $\chi(\lambda)$. For the gaussian case considered above the change in the entropy due to this restriction is inessential (to be published).

In summary, we have shown that the entanglement entropy can be directly inferred from statistics of current fluctuations. We derived a general relation between the entanglement entropy and electron transport via the full counting statitics. Builiding on this universal relation, we propose noise measurement in a QPC as a way to test theoretical predictions for the many-body entanglement in a realistic setting. This provides a new method to invetstigate many-body entanglement, and in particular, its generation in non-equilibrium quantum systems.



The work of I.K. was supported in part by the National Science Foundation under Grant No. PHY05-51164. 
L.L.'s work was partially supported by W. M. Keck foundation. We thank Carlo Beenakker and Gil Refael for comments on the manuscript and useful discussions.


\begin{thebibliography}{99}
\bibitem{Bernoulli_numbers}
The Bernoulli numbers $B_{m}$ are defined by the generating function $\frac{x}{e^x-1}=\sum_{n=0}^{\infty}B_n {x^n\over n!}$.   
Asymptotically, $|B_{n}|\approx{2\, n!\over (2\pi)^{n}}$ for large even $n$,
which shows that the coefficients in (\ref{Entropy from cumulants}) 
stay bounded for high order cumulants.
\end{thebibliography}
\end{document}